# Efficiency of Covid-19 Mobile Contact Tracing Containment by Measuring Time Dependent Doubling Time

Antonio Bianconi [1,2,3], Augusto Marcelli [1,4], Gaetano Campi [1,2] and Andrea Perali [1,5]

[1] Rome International Centre Materials Science Superstripes RICMASS via dei Sabelli 119A, 00185 Rome, Italy; antonio.bianconi@ricmass.eu

[2] Institute of Crystallography, CNR, via Salaria Km 29. 300, Monterotondo Stazione, Roma I-00016, Italy; gaetano.campi@ic.cnr.it

[3] National Research Nuclear University MEPhI (Moscow Engineering Physics Institute), 115409 Moscow, Russia

[4] INFN - Laboratori Nazionali di Frascati, 00044 Frascati (RM), Italy; augusto.marcelli@lnf.infn.it

[5] School of Pharmacy, Physics Unit, University of Camerino, 62032 Camerino (MC), Italy. andrea.perali@unicam.it

\* Corresponding author: antonio.bianconi@ricmass.eu; tel. +39 3388438281

ORCID; Antonio Bianconi Augusto Marcelli 0000-0002-8138-7547

Gaetano Campi 0000-0001-9845-9394 Andrea Perali 0000-0002-4914-4975

*Abstract*

The Covid-19 epidemic of the novel coronavirus (severe acute respiratory syndrome SARS -CoV-2) has been spreading around the world. While different containment policies using non-pharmaceutical interventions have been applied, their efficiency are not known quantitatively. We show that the doubling time $T_d(t)$ with the success *s factor*, the characteristic time of the exponential growth of $T_d(t)$ in the arrested regime, is a reliable tool for early predictions of epidemic spread time evolution and it provides a quantitative measure of the success of different containment measures. The efficiency of the containment policy *Lockdown case Finding mobile Tracing* (LFT) using mandatory mobile contact tracing is much higher than the *Lockdown Stop and Go* (LSG) policy proposed by the Imperial College team in London. A very low *s* factor was reached by LFT policy giving the shortest time width of the dome of positive case curve and the lowest number of fatalities. The LFT policy has been able to reduce by a factor 100 the number of fatalities in the first 100 days of the Covid-19 epidemic, to reduce the time width of the Covid-19 pandemic dome by a factor 2.5 and to rapidly stop new outbreaks avoiding the second wave.

## 1. Introduction

The diffusion of Covid-19 is a transnational phenomenon that involves all continents reaching a million of positive cases and thousands of deaths at early April, 2020. Following the threshold of the very fast Covid-19 epidemic in January 2020 in Wuhan [1,2], it was found that a very small characteristic time (about 2 days) of the exponential growth and the characteristic number $R_0$ of infected humans by one infected case was about $R_0$=2.3, much larger than the critical value of 1, pointing to an explosion of the pandemic with possible millions of infected people in few weeks [3].

In the absence of a Covid-19 vaccine and without established therapies, scientists informed the policy makers for the urgent need of actions for epidemic containment measures [4,5] to reduce the expected number of hundreds of thousands of deaths in the short time of the pandemic peak. Non-pharmaceutical interventions to reduce the care need requests, below the critical number of beds available in hospitals, have been proposed for USA [6,7] based on the experience on SARS containment invoking in addition fast containment measures such as self-isolation. The lockdown stop and go protocol has been proposed by Ferguson group of Imperial College [8] based on simulations with models developed on the basis of known influenza pandemic [9,10]. The epidemic control measures look for the reduction of the number of daily new cases *N(t)* to avoid an unacceptable load to the health care systems to reach the goal of reducing the number of deaths. The traditional method to reduce the number of deaths was the lockdown, an emergency protocol that is enforced to prevent people from leaving a given area, city or a region. This containment measure expands the time lapse of the virus diffusion by stretching the exponent of its exponential growth. Therefore, while the number of cases will decrease at the top, the time extent of the epidemic will become longer, however with a severe negative impact to the economy.

China first focused an unconventional Covid-19 policy to reduce both the number of deaths and the time extent of the epidemic [1]. This unconventional policy called here *Lockdown, case Finding, mobile Tracing* (LFT) was based on the combination of measures looked to reduce both the number of new daily cases at epidemic peak and to reduce the time width of the epidemic dome. Achieving the reduction of both maximum and width of the diffusion peak, China obtained the drastic reduction of the total number of deaths. After about 80 days from the epidemic threshold, on April 7$^{th}$ 2020 the long-term lockdown was *stopped* in China with officially less the 3400 fatalities. It has been the largest "experiment" to test the efficiency of the new LFT policy in the history of the epidemiology, which takes advantage of both mass search of positive cases and the tracing of infected cases by a mobile phone application. A similar approach has been considered by other countries, all taking advantage of advanced methods based on Apps, i.e., software applications designed to run on mobile phones and the treatments of Big Data developed in the last years. The success of the Chinese and south Korean Covid-19 policy was considered by other countries such as Norway, Singapore, and Taiwan [5].

The more conventional approach - a type of "*herd immunity*" - has been considered by European countries, and USA, called *Lockdown stop and go* (LSC), [8]. The LSC approach is made of combination of advices for the population to keep "*physical distancing*" to protect others, to stay at home only for positive cases, household quarantine of their family members and to reduce traveling. Intermittent measures are planned to be temporarily relaxed in short time windows, and reintroduced when positive case numbers rebound [8]. The media informed the population about the actual numbers of epidemic cases and the advices were diffused only weeks after the Covid-19 pandemic onset.

Italy, France and Spain, have followed the LSG policy and in March 2020 used the mitigation strategy called here *Mandatory Full Lockdown* (MFL) with few days delay, where the LSG traditional epidemiologic confinement approach was strongly enforced. These unprecedented measures of imposition by law all over the country consist of physical distancing, ordering enforced closure of schools, universities, all national manufactures, ban of mass gatherings and public events, and confinement at home of the entire population. This approach had the key target of reducing the number of infected cases day per day paying attention not to overcome the maximum number of sick person requiring critical care [6-8]. In fact, the MLF policy gives priority to the health care system respect to other economic demands.

Academic epidemiology analysis is usually made at the conclusion of the epidemic event. On the contrary, in the case of the Covid-19 epidemic some countries immediately shared verified data and made them available in public repositories. This opportunity gives scientists the possibility to shed light on the new physics of epidemics with containment measures [11-19] a growing field of high interest for population, health systems [8.9], travels [20,21] and economy [10]. This is a highly interdisciplinary research activity, at the intersections among biological physics, advanced statistical physics of complex systems, physics of quantum complex materials, chain reactions control in nuclear reactors and the most recent *Big Data* analysis methods.

It was clear for all scientists that in the early days Covid-19 follows the classical exponential law of epidemic spreading with a constant rate, but it was realized by few [11-19] that it is followed by a second regime with a variable time dependent rate quite different for countries applying different policies. However, in spite of the relevance of the question, none was able to measure in a quantitative way the relative efficiency of the different containment measures based on verified Covid-19 data released by official institutions and health agencies, a key information need to stop the diffusion of this world-wide epidemic.

Only two weeks after the start of the MFL policy, it was [13,14] pointed out that the slowing down of the pandemic diffusion was much less effective for the MLF that for the LFT approach. The data analysis approach [13,14] used the quantitative determination of the time dependent doubling time $T_d(t)$ of the Covid-19 pandemic spread calculated averaging *day by day* data over a 5 days interval. The measure of the time dependent doubling time ($T_d$) is widely used not only in epidemiology, but also in the nuclear reactor physics where it is used to keep the reactor in the critical regime between the subcritical regime of the chain reactions of nuclear fission processes and the supercritical regime with risks of severe explosive accidents. The similarity of the nuclear chain reactions in uranium with biological cell fission was already noticed by Lise Meitner and Otto Robert Frisch when they coined the name nuclear fission, in 1939.

The comparative analysis of the time dependent doubling time data in South Korea and Italy on March 15[th] after about 20 days from the epidemic threshold $t_0$ has quickly shown that the LFT controlled Covid-19 epidemic growth entered in the arrested phase in South Korea while in Italy it was still rapidly growing in the near threshold phase [13,14]. This was possible because the doubling time analysis has shown two well separated regimes: the first near threshold phase, described by a stretching exponential with a slowly increasing stretched characteristic time, followed by a second phase, the arrested growth process following the Ostwald growth over the course of time, where one phase transforms into another metastable phase, but with a similar free energy [13,14,22-24]. This mechanism has been observed in the diffusion of oxygen interstitials diffusion in quantum complex matter [25-27] and in the crystallization of complex molecules [28] and proteins [29].

## 2. Results and Discussion

The different efficacy of the containment policies can be recognized in Figure 1a where the cumulative number of cases $N_c(t)$ in different countries deposited on the data banks on the 95[th] day of the year, is plotted in the time scale with the zero sets at the first day $t_0$ of the exponential growth. The time $t_0$ of the epidemic threshold has been defined as the day where the time dependent doubling time shows a minimum before it starts to increase. Using this criterion the cumulative number of cases shown in Fig.1a follows an average exponential law from the 5th to the 14th day with a doubling time about 3 days and about 60 cases at the initial time $t_0$ of the epidemic threshold.

As it can be seen in the panel (a) of Figure 1 the diffusion rate for the different policies in various countries is similar in the near threshold regime. Indeed, the reported curves of the cumulative number of cases $N_c(T)$ overlap in the near threshold regime during the first days while they strongly diverge in the arrested regime. The dome of the positive case curve in USA and UK has been predicted [7,8] using the standard individual-based simulation model developed to support pandemic influenza planning [9,10] are plotted in panels (b) and (c).

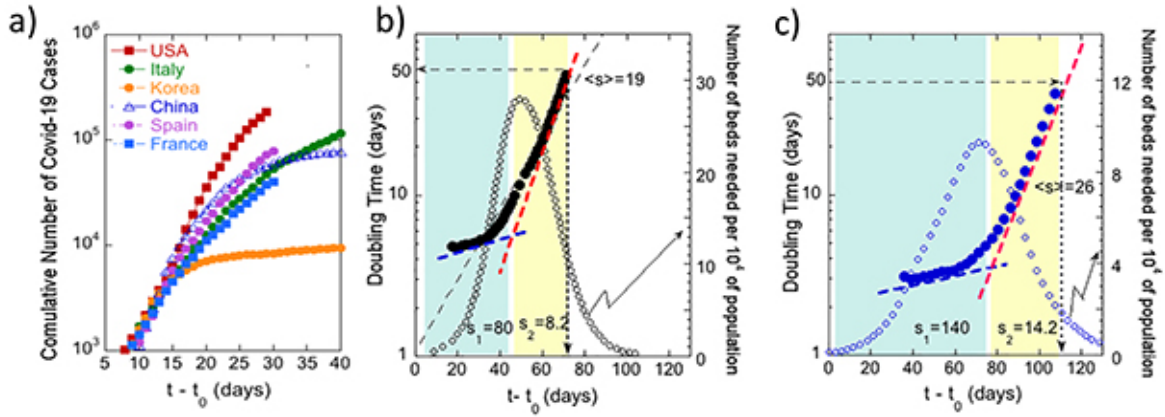

**Figure 1.** Standard curves used to report the Covid-19 epidemic time evolution showing the cumulative number of cases (panel a) and the standard theoretical epidemiology predictions of the needed number of beds based on the prediction of daily new cases for an uncontrolled epidemic case (panel b) and mitigated epidemic case (panel c). ***Panel a)*** Cumulative number of positive cases $N_c(t)$ vs. time in South Korea (orange), France (blue), China (light blue), Italy (green), Spain (magenta) and USA (red). The time scale for each curve of each country starts on the threshold day $t_0$. The curves overlap in the near threshold regime while later separate. The curves of the cumulative number of cases of China and South Korea become flat after about 30 days, when the pandemic is arrested. ***Panel b) & c)*** the theoretical predictions for the number of needed intensive care beds (open symbols) from [8] for an uncontrolled diffusion (**b**) and for severe containment measures (**c**). We show in Panel b) our calculation of the exponential variation of the doubling time $T_d(t)$. The time dependent doubling time curves calculated in this work for uncontrolled (black curve in panel b) and controlled epidemic (blue curve in Panel c). Our calculations predict a different exponential increase of the doubling time in the near threshold regime (blue area) and in the arrested regime (yellow area). The stop of the dome of the positive case curve occurs in the range 50<$T_d(t)$<70 d where the reproduction number R(t) becomes less than one, entering in the critical regime where the effective infection time becomes equal to the effective removal time.

The calculated time dependent number of hospital beds dedicated to intensive care medicine for the case of wild uncontrolled epidemic in panel (b) will imply the failure of both USA and UK health care systems [6-8]. To avoid the break down it was proposed to reduce the peak intensity in panel (b) by expanding the epidemic time scale by a factor 1.5 using non-pharmaceutical interventions, e.g., case isolation, home quarantine and physical distancing, as evident in panel (c). In this work in order to evaluate the time evolution of the epidemic growth, we have used the time dependent doubling time $T_d(t)$ as the key physical parameter,

$$T_d(t) = \frac{\ln(2)}{\frac{d[\ln(N_c(t))]}{dt}} \qquad (1)$$

where $N_c(t)$ is the cumulative number of cases and the derivative at each time $t$ is obtained by averaging the $N_c(t)$ curve over a period of five days.

The efficacy of the containment policies is probed by the increase of $T_d(t)$ from its minimum value $T_{d0}=2$ days, at the time threshold $t_0$ to the time t* where $T_d=50$ days. This value of the doubling time is where the exponential growth in the supercritical phase ($T_d$ <50) is expected to stop because it is the lowest limit of the Covid-19 epidemic critical phase, in the range 50<$T_d$<100 days, where the reproduction number $R_0$ is near one [30]. Therefore, we have calculated the curves $T_d(t)$ in panels (b) and (c) of Figure 1, which shows the exponential increase of $T_d(t)$ in the time range of the exponential growth showing the dome of the positive case curve [13,14]. A kink in $T_d(t)$ separates two exponential increasing regimes: the near threshold regime (shaded blue region) and the arrested regime (shaded yellow region) separated by the transition regime around the peak of the pandemic curve $N(t)$ of the number of new daily cases. The theoretical curves of $T_d(t)$ for the uncontrolled epidemic (black filled dots in panel b) and with the lockdown mitigation (blue filled dots) show relevant changes on the time evolution behavior, which depends on the containment policy. First, in the near threshold regime, the doubling time follows an exponential growth (blue line in the semi-log scale) with the characteristic time $s_1$

$$T_{d1} = C_1 e^{t/s1} \qquad (2)$$

and in the arrested regime it follows a second exponential growth (red line in the semi-log scale) with the characteristic time $s_2$

$$T_{d2} = C_2 e^{t/s2} \qquad (3)$$

The figure shows that the theory predicts that $s_2$ is much smaller than the characteristic time $s_1$. In fact, measuring the average $<s>$ factor by fitting the full $T_d(t)$ curve in the range $2 < T_d < 50$ [13,14] we have provided a quantitative measure of the efficiency and the effectiveness in term of time of the enforced containment policy. The extraction of $<s>$, $s_1$, and $s_2$ factors allows a straightforward quantitative evaluation and, in addition, a quantitative comparison of the different containment policies adopted to control the epidemic become possible. The calculations show that the factor $s_1$ decreases in the near threshold regime from the wild regime value $s_1=80$ days in panel (b) to the lockdown regime $s_1=140$ days in panel (c). The factor $s_2$ in the arrested regime increases from the wild regime value $s_2=8.2$ days in panel (b) to the lockdown regime $s_2=14.2$ days in panel (c).

The doubling time $T_d(t)$ reaches the value of 50 days after more than two months in the wild uncontrolled pandemic spread curve, while this value is reached after 110 days in the most severe lockdown policy, i.e., the pandemic is predicted to be about 1.6 times longer in the lockdown regime, in agreement with the ratio of the half width values at half maximum.

The explosion of the Covid-19 pandemic has seen a fast response of the scientific community proposing different models, which can be quickly verified or falsified analyzing experimental data by fast computer codes, useful to test the different containment measures [11-19].

The key results of our work are shown in Figure 2 and 3 where we report the doubling time $T_d(t)$ as a function of time, extracted by verified data *vs.* time with the zero sets at the threshold time $t_0$ in several countries where different Covid-19 epidemic policies have been enforced. The time evolution of the experimental doubling time $T_d(t)$ for South Korea and China where the *Lockdown, case Finding, mobile Tracing* (LFT) policy was applied is plotted in panel (a). The dome of the positive case curve in both countries is shown by the curve of verified data of the number $N(t)$ of daily new cases.

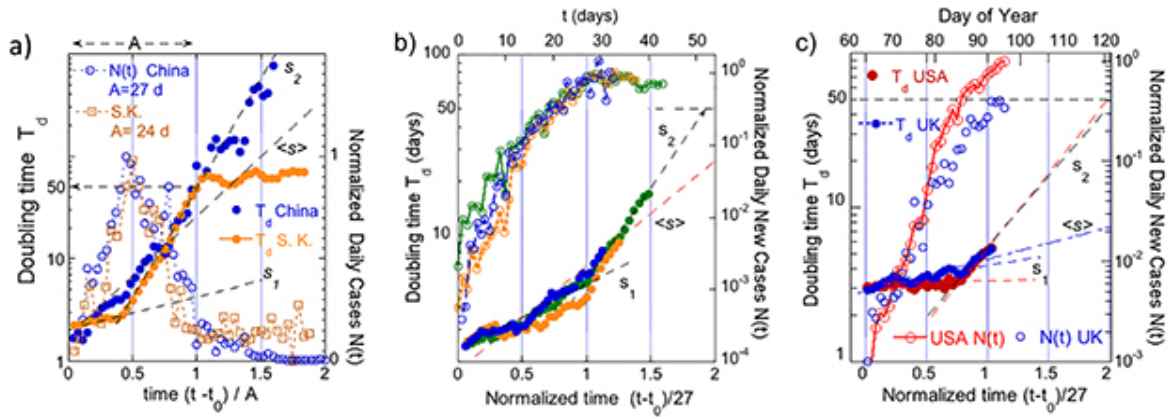

**Figure 2.** In this figure we report the time evolution of the doubling time $T_d$, which we propose here to use to shed light on the unconventional time evolution of the Covid-19 pandemic with containment measures. **Panel a)** Evolution of the experimental doubling time $T_d(t)$ in the countries where the LFT policy was enforced: South Korea (filled orange dots) and China (filled blue dots). The open circles are the known curves of the number of daily new cases $N(t)$ which are used here to track the time evolution pandemic domes. **Panel b)** the experimental time dependent doubling time $T_d(t)$ in Spain (filled orange dots), France (filled blue dots) and Italy, (filled blue dots) where the *Mandatory Full Lockdown* MFL policy was applied. **Panel c)** the evolution of the doubling time in USA (filled red dots) and UK (filled blue dots) applying the policy called *Lockdown stop and go* (LSC). The $N(t)$ curves are plotted in semi-log scale in panel b and c. In panel (a) and (b) the numbers in the $N(t)$ curves are normalized to overlap all curves in the near threshold region. The near threshold time regime in panel (b) and panel (c) identified in the range between the threshold and the kink in the $T_d(t)$ is 27 days in panel (b) and similar in panel (c).

The experimental doubling time $T_d(t)$ where the time unit is one day, is calculated using equation (1) and it does not need any normalization. In South Korea, after the threshold of epidemic outbreak, $T_d(t)$ increases exponentially up to the time t* where $T_d(t)$ reaches the value of 50 days. For t > t*, $T_d(t)$ remains constant showing the transition from exponential to linear growth. In our investigation of the Covid-19 outbreak in Italian regions [30] we have extracted both the time dependent reproduction number R(t) and the doubling time $T_d(t)$ showing that the critical phase where, $R_t \approx 1$ correspond with the doubling time fluctuating in the range $50<T_d(t)<100$. This regime separates the exponential growth phase (which is called supercritical), with $R_t >1$, from the phase with $R_t<1$ (which is called subcritical). Therefore, the time width of the exponential growth regime giving the characteristic width of the dome of the Covid-19 positive case curve, can be measured by the difference between the day T*, where $T_d$ reaches the value of 50 days, and the outbreak threshold time $t_0$. We have found the full time width of the epidemic dome to be 27 days for China and 24 days for South Korea. Introducing a normalized scale, i.e., dividing the time in the time axis by the full time width, both the normalized domes of $N(t)$ and the experimental $T_d(t)$ curves for China and South Korea fully overlap, providing evidence for the characteristic behavior associated with the mandatory "contact tracing" LFT policy.

The key result of the data analysis is that the kink in the $T_d(t)$ curves, which separates the near threshold regime from the arrested regime is only 13-14 days. The different value of the $s_1$ factor in the near threshold regime is due to the immediate activation of the LFT policy with mandatory mobile contact tracing as requested by scientists and experts, [3,4] which was adopted by policy makers in South Korea faster than in China.

The overlapping domes of the positive case curves in both countries shows also that both the epidemic peak and its full time width at half maximum are strongly reduced by the LFT policy with mandatory "contact tracing". In Table 1 we report the key parameters obtained by the fits to unveil the physics of the time evolution of pandemic in the selected countries.

Panel (b) of Figure 2 shows that the time dependent doubling time $T_d(t)$ in Italy, Spain and France, where *Mandatory Full Lockdown* (MFL) policy (without the mandatory "contact tracing" LFT policy) was applied, fully overlap. In the near-threshold regime the values of the doubling time are very similar and the three curves show the same kink, probing the transition from the near-threshold to the arrested regime at the same moment, i.e., 27 days, which has been used to normalize the time scale. On 102[th] day of the year (DoY) the arrested regime in these countries was only at its early stage, but it shows similar *s* factors.

The time dependent doubling time for USA and UK that adopted later a more soft version of the *Lockdown stop and go* LSG policy is plotted in panel (c) of Fig. 2. While scientists and some policymakers asked their governments to activate as soon as possible containment measures [5], but for some time policymakers of USA and UK have chosen to "do nothing". The effect has a clearly impact on the $s_1$ factor as outlined by the flat $T_d(t)$ curve in the

Antonio Bianconi et al. *Efficiency of Covid-19 Containment by Measuring Time Dependent Doubling Time*

near-threshold regime showing a similar constant behavior for USA and UK with a very large $s_1$ factor.

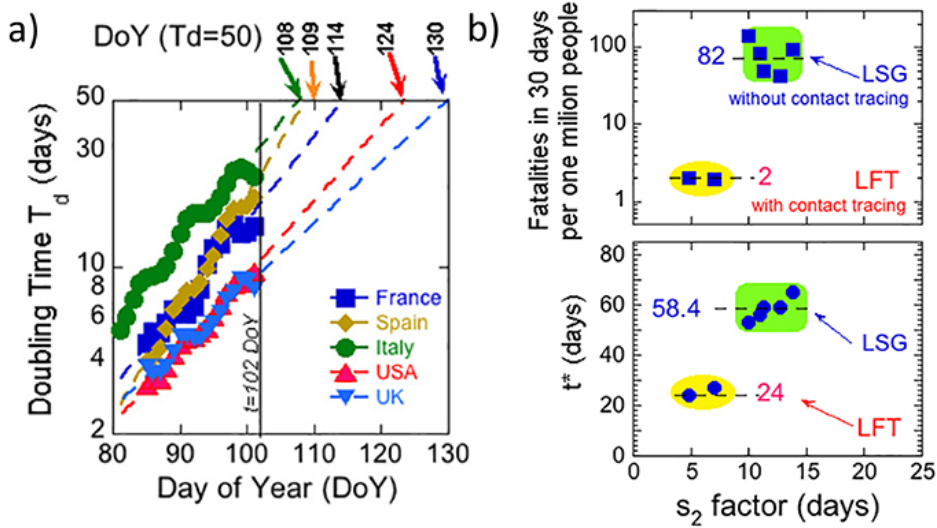

**Figure 3.** Extrapolated time T* of the day of year (DoY) when the doubling time $T_d$ will reach 50 days $T_d(T^*)=50$, used to predict the time width of the supercritical epidemic regime t*=T*-$t_0$ for LSG countries, from data available on the $102^{th}$ DoY. The quantitative test of the efficiency of one containment policy versus another one is given by its ability i) to achieve a short time width t* and ii) to reduce the number of fatalities in the same time period normalized to the population. *Panel a)* Plot of the doubling time $T_d(t)$ as a function of time in the arrested regime for the epidemic spread in Italy, France, Spain, UK and USA. The linear fit in the semi-log scale allows the prediction of the expected day T* of the year when $T_d$ will be 50. The predicted time duration of the supercritical regime is given by t*=T*-$t_0$. *Panel b)* Plot of t* as a function of the success $s_2$ factor. The lower panel shows that both the $s_2$ factor and t* in the LFT countries (data in the yellow area) are smaller than in LSG countries (data in the blue area). The average value t*=24.8 days in countries where the LFT protocol has been enforced is about a factor 2 lower than the average value t*=58.4 days in countries where LSG protocol [8] has been followed in agreement with data in Fig. 2b. The upper panel (b) shows that the normalized number of fatalities in 30 days per milion people in the countries which have adopted the LSG protocol without mandatory mobile contact tracing are similar and their average, $N_{30}$=82, it is a factor 40 larger than the number $N_{30}$=2 reached in the countries which have adopted the LFT protocol with mandatory mobile contact tracing.

| Data of 102 DoY | a) $t_0$ outbreak time threshold (DoY) | b) $T_{d3}$ at 3rd day | c) $s_1$ (days) | d) $s_2$ (days) | e) Predicted $T_s^*$ (DoY) when $t_d=50$ is expected | f) Predicted $t^*=T^*-t_0$ time interval of the supercritical phase (days) | g) Population (millions) | h) Fatalities in the first 30 days per million people |
|---|---|---|---|---|---|---|---|---|
| USA | 65±1 | 3±0.2 | 388 | 12.7±1 | 124 ± 2 | 59±4 | 328.200 | 42.563 |
| Italy | 52±1 | 2.5±0.2 | 30 | 11±1 | 108 ± 2 | 56±4 | 60.360 | 84.601 |
| Spain | 56±1 | 2.7±0.2 | 56.5 | 10±1 | 109 ± 2 | 53±4 | 49.940 | 139.03 |
| France | 55±1 | 2.6±0.2 | 29.5 | 11.3±1 | 114 ± 2 | 59±4 | 66.990 | 49.764 |
| UK | 65±1 | 3±0.2 | 84 | 13.8±1 | 130 ± 2 | 65±4 | 66.650 | 94.432 |
| China | 22±1 | 2±0.2 | 9.4 | 7±1 | 49±1 | 27±2 | 1393.000 | 1.9093 |
| South Korea | 48±1 | 2.3±0.2 | 43 | 4.8±1 | 72±1 | 24±2 | 51.640 | 2.0365 |

**Table 1**: The parameters of the time evolution of Covid-19 pandemic in different countries calculated from data available on the 102$^{th}$ Day of the Year 2020. **a)** $t_0$, DoY of the threshold of epidemic national outbreak; **b)** doubling time $T_d$(3d) at the third day from the threshold; **c)** $s_1$ i.e., the *s*-factor in the near threshold regime; **d)** $s_2$, i.e., the *s*-factor in the arrested regime; **e)** Prediction of the Day of the Year T* when the doubling time $T_d(T^*)$ will reach the value of 50 days. **f)** Prediction of the time duration t*= T*-$t_0$ of the Covid-19 epidemic exponential growth in the supercritical regime with doubling time $T_d(t^*)$ <50*d*; **g)** population of each considered country; **h)** number of fatalities per million people during the first 30 days of the national outbreak in each country.

We have plotted in panel (a) of Figure 3 the exponential increase of the doubling time in the arrested regime for all 5 investigated countries applying the LSG policy without the mandatory "contact tracing" policy on 102$^{th}$ day of the year. From the extrapolation of the line in the semi-log scale (as determined by the exponential curve of $T_d(t)$ in the arrested regime) it is possible to predict the time duration of the supercritical phase showing the exponential growth in all five countries which adopted the LSG policy.

The exponential growth is expected to enter in the critical regime when the doubling time $T_d$ (t*) > 50 days, as shown for the case of South Korea in Fig. 1. However, it is not possible to predict how long this critical regime will continue since the number of daily new cases could show either large fluctuations, or could decrease or increase with a linear rate, as it has occurred in South Korea for a long time.

The linear fit of the doubling time in the arrested regime $T_d(t)$ in the semi-log scale shown in panel (a) of Fig. 3 for the five cases where the LSG policy was enforced: Italy, France, Spain, UK and USA, allows us to predict the expected day T* of the year when $T_d$ will be 50 (called DoY stop): 108 in Italy, 109 in Spain, 114 in France, 124 USA and 130 in UK with the error bar of ~2 days. The time width t* of the epidemic dome in the positive case curve in the LSG countries is given by the time difference $t^*=T^*- t_0$, which is predicted to be in the range 55-63 days as given in table 1, i.e., it is more than two times longer for countries where the mandatory "contact tracing" LFT policy was not adopted. In Table 1 we have summarized the key numerical results based on data available on April 10$^{th}$.

In lower panel of Fig. 3b we report the estimated time width t* of the Covid-19 epidemic supercritical regime as a function of the $s_2$ factor. The success $s_2$ factor in countries which enforced the LFT policy with mandatory "contact tracing" has been found to be in the range $5 < s_2 < 7$ while the $s_2$ factor in countries which enforced the LFT policy without mandatory "contact tracing" falls in the range ($10 < s_2 < 15$).

The time width t* in the countries which have applied the LSG protocol proposed by the group of Imperial College [8] without mandatory contact tracing is predicted to cluster around the average value of 58.4 days in the blue area which is two times larger that the average value 24.8 days in countries where the LFT protocol has been enforced in agreement with the differences of the time width of the epidemic dome of the positive case curve shown in Fig.2.

In order to evaluate quantitatively the efficiency of the different policies for taking care of the health of the population in different countries it is necessary to calculate the number of fatalities collected in the same time interval and normalized to the total number of country population. On the 102 DoY we can analize all cumulative curves of the Covid-19 fatalities of all studied countries in the same time interval covering only the first 30 days of the epidemic spreading. In the upper panel of Fig. 3b we have plotted the fatalities in the first 30 days, divided by the number of inhabitants in different countries measured in millions as a function of the $s_2$ factor. The data from countries which have applied the Lockdown case Finding mobile Tracing (LFT) policy with mandatory mobile tracking [13-15] occur in the yellow region showing that they have achieved a both a lower $s_2$ factor and a low number of fatalities. The countries in the blue region, using the Lockdown Stop and Go (LSG) protocol [8] with no mandatory mobile tracking, have achieved a $s_2$ factor about two times longer. Morever on April 10 the average number of normalized fatalities is 40 times higher in all countries using the LSG protocol without mandatory contact tracing than in countries usong the LFT protocol with mandatory contact tracing.

## 3. Validation of predictions and evaluation of containtment protocols

On June 25th 2020, in the course the peer review process of this paper, we have been invited to check our results by analysis of new available data of the Covid-19 epidemic spreading. The available data on the 177$^{th}$ DoY of 2020 allow us to make a comparative evaluation of the efficiency of both LFT and LSG containtment protocols in the same time range, 100 days, covering the full dome of the positive case curve and to validate the accuracy of predictions based on data covering only the first 15 days or the first 30 days [13,14]. The time evolution of the doubling time in the first 100 days from the outbreak time threshold are shown in panel (a) of Fig. 4 for countries folowing the LFT protocol and in panel (b) of Fig. 4 for countries folowing the LSG protocol. The supercritical phase, characterized by the exponential growth of epidemic following the explosion of the outbreak, which occurs for t<t* where $T_d(t)$<50d [30] is indicated by the red area. The the yellow area, where 50 d <$T_d(t)$< 100 d, indicates the critical regime. The time duration of the supercritical phase is much shorter, i.e., 24<t*<27 days as shown in panel (a) for LFT countries, than in LSG countries, which is 52<t*<78 days as shown in panel (b). The time width of the supercritical regime has been much longer than predicted in Fig. 3 for UK and USA. The success $s_2$ factor, measured in the supercritical regime for the LSG countries, is in the range 11<$s_2$<24 (in the blue area of panel b). It is larger than predicted in Fig. 3 and it is much higher than in the LFT countries, in the range 5<$s_2$<7 (panel a).

On the 177 DoY it is possible to analize the cumulative curves of the Covid-19 fatalities of all studied countries in the same time interval of 100 days, covering the full width of the epidemic wave. The impact of different containment policies on the population has been measured by the relation between the number of fatalities in the first 100 days of Covid-19 epidemic spread per milion people ($N_{100}$) and the $s_2$ factor shown in panel (c) of Fig. 4. The $s_2$ factors and the number of fatalities $N_{100}$ are reported in Table 2.

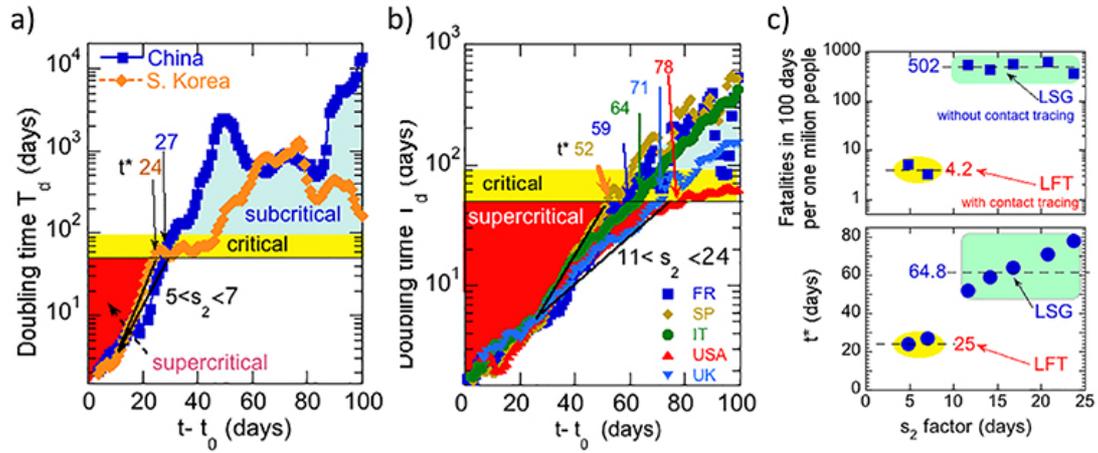

**Figure 4.** The analysis of data available on the 177[th] DoY have allowed us to measure the time dependent doubling time $T_d(t)$ in the first 100 days after explosion of the national outbreaks covering the time full width of the epidemic dome in the supercritical regime. *Panel* **a)** shows the curves $T_d(t)$ in countries which adopted the LFT protocol with mandatory mobile contact tracing. *Panel* **b)** shows the curves $T_d(t)$ in countries which adopted the LSG protocol without mandatory mobile contact tracing. The supercritical regime indicated by the red area, characterized by the exponential growth of Covid-19 epidemic following the explosion of the outbreak occurs for t<t* where $T_d(t)<50d$. The time t* where $T_d(t^*)=50d$ is assigned to the transition from the supercritical regime to the critical regime (yellow area) which occurs in the range $50d<T_d(t)<100d$. The subcritical regime (blue area) occurs where $T_d(t)>100d$. The success $s_2$ factor measured in the supercritical regime is in the range $11<s_2<24$ in the LSG countries (panel b) while is much smaller in the range $5<s_2<7$ (panel a) for the LFT countries. The duration t* of the supercritical regime is in the range 52<t*<78 days, with the average value <t*>=64.8 days, in LSG countries, (panel b) to be compared with 24<t*<27 days in LFT countries. *Panel* **c)** shows the time duration t* of the supercritical regime (lower panel) and the number of fatalities in the first 100 days per million people (upper panel) as a function of the $s_2$ factor. The lower panel shows that both the $s_2$ factor and the t* in the LFT countries (yellow area) are shorter than in LSG countries (blue area). The upper panel shows that the normalized average number of fatalities in 100 days over the full dome of the epidemics is about 129 times higher in countries where LSG protocol [8] has been enforced ($N_{100}$=502) than in countries where the LFT protocol has been enforced ($N_{100}$=42).

Qualitative predictions made in and predictions made on April 10 (see panel (b) in Fig. 3) are confirmed. The upper panel of Fig. 4c shows the number of fatalities ($N_{100}$) in the first 100 days per million people as a function of the $s_2$. The results based on data collected on the June 25th in panel (c) of Fig.4 confirm qualitatiely the predictions made on March 15 [14-15] and April 10th which are shown in Panel (c) of Fig. 3.

| Data collected on 177 DoY | a) $s_2$ (days) | b) $T^*$ measured time (DoY) when $t_d=50$ | c) measured $t^*=T^*-t_0$ time interval of the supercritical phase (days) | d) Fatalities in the first 100 days per million people |
|---|---|---|---|---|
| USA | 23.7±1 | 143±1 | *78±2* | 357.55 |
| Italy | 16.8±1 | 116±1 | *64±2* | 552.60 |
| Spain | 11.6±1 | 108±1 | *52±2* | 543.37 |
| France | 14.1±1 | 114±1 | *59±2* | 436.55 |
| UK | 20.7±1 | 136±1 | *71±2* | 623.72 |
| China | 7±1 | 49±1 | *27±2* | 3.3288 |
| South Korea | 4,8±1 | 72±1 | *24±2* | 5.2285 |

**Table 2**: The parameters of the time evolution of Covid-19 pandemic in different countries calculated from data available on the 177[th] Day of Year 2020. **a)** the $s_2$ factor of the exponential growth of the doubling time in the arrested regime of the supercritical phase; **b)** The measured Day of the Year $T^*$ when the doubling time $T_d(T^*)$ has reached the value of 50 days; $T_d(T^*)=50$ *d*; **c)** the measured time interval $t^*=T^*-t_0$ of the supercritical phase with the epidemic exponential growth rate where $T_d(t)<50$ *d*; **d)** number of fatalities per million people (normalized to the country population) during the first 100 days of the national outbreak of each country.

All results confirm that a lower s-factor is correlated with both short duration of the supercritical epidemic regimne and a lower number of fatalities. At the end of the first pandemic wave the duration of the supercritical phase has been between 2 and 3 times longer using the LSG protocol than using the LFT protocol. The normalized number of fatalities has been about 120 higher in countries that applied the LSG protocol proposed by the Ferguson team at the Imperial College [8] than in countries that applied the LFT protocol with mandatory contact tracing. The LSG protocol guarantees privacy, but the data show that it provides a poor public health care since the number of Covid-19 fatalities has been a factor 120 times higher using the LSG protocol than the LFT protocol using mandatory mobile contact tracing.

## 4. Conclusions

In this work we have extracted the time evolution of both the doubling time $T_d(t)$ and the success s-factor, by analysis of available data of Covid-19 epidemic in seven countries where different containment policies have been applied. The key result of this work has been to unveil the presence in the same dome of the positive case curve of two statistical different regimes during the exponential growth: i) the stretched near threshold growth phase and ii) the arrested phase.

The two phases show two exponential functions of $T_d(t)$ versus time separated by a kink, characterized by two time exponents $s_1$ and $s_2$. These success *s*-factors are used to quantify the efficacy and success of the different mitigation methods. They could and should be used in the future to monitor with accuracy and in a quick way, possible rebounds of this pandemic, but it is also valid for a future pandemic (if any).

We clearly show that countries that selected and adopted advanced technologies [13,14,15] i.e., the containment policy *Lockdown, case Finding, mobile Tracing* (LFT) with mandatory "mobile contact tracing" have been able to reduce both the peak and the width of the *epidemic dome* of the daily positive case curve of the Covid-19. The reduction of the time duration of the lockdown as obtained by mandatory "contact tracing" has minimized the impact on the economy [10] keeping the manufacture close for a shorter time. The number of fatalities per million people over 100 days, covering the full width of the first wave, has been more than a factor 100 smaller than in the countries, which have not used mandatory mobile contact tracing.

Finally, we have shown in this work that the proposed use [13,14,31] of the time dependent doubling time plots for early warning on the epidemic spread rate gives reliable predictions in particular in the early stage of the epidemic spreading. Therefore, the proposed method can be considered useful to predict the time evolution of future epidemic outbreaks. This approach represents a fundamental advantage compared to the standard analysis of the Covid-19 epidemic and supports the theory of fractal statistics or new complex statistics of epidemic spread [12,19]. In this study, based on data published on the 102[th] day of the year for the epidemic growth in action in Italy, France, Spain, UK and USA we predicted (see panel (c) in Fig. 3) the end of the dome of the positive case curve for each country on the day of the year reported in Table 1 within the indeterminacy of a couple of days. The results have been validated in Figure 4 and Table 2 covering the first 100 days of the Covid-19 epidemic in different countries. All predictions have been validated by data available on the 177[th] DoY and a quantitative comparative valuation of different protocols used for Covid-19 containment have been obtained. Finally, we hope that this original approach will be of help to understand epidemiology of Covid-19 and for policymakers that are now well informed [31] how to save lives and reduce fatality numbers by two orders of magnitude in a second wave of Covid-19 or for a future pandemic.

**Acknowledgments**

We acknowledge the non-profit organization *Superstripes-onlus* for funding this research.

***Author Contributions:*** The authors contributed equally to the conceptualization, methodology, and investigation while preparing the article.

***Conflicts of Interest*:** The authors declare no conflicts of interest.